\documentclass[twocolumn,showpacs,preprintnumbers,amsmath,amssymb]{revtex4}

\usepackage{graphicx}
\usepackage{dcolumn}
\usepackage{bm}

\newcommand{\ybco}{YBa$_2$Cu$_3$O$_{7-\delta}$ }

\begin{document}


\title{Pseudogap state in slightly doped by aluminium and praseodymium YBa$_2$Cu$_3$O$_{7-\delta}$ single crystals with a given topology of plane defects}

\author{R. V. Vovk}
 \email{Ruslan.V.Vovk@univer.kharkov.ua}
\author{M. A. Obolenskii}
\author{A. A. Zavgorodniy}
\author{D. A. Lotnyk}
\author{K. A. Kotvitskaya}
\affiliation{Physical department, V.N. Karazin Kharkov National
University, 4 Svoboda Square, 61077 Kharkov, Ukraine.
}%

\date{\today}

\begin{abstract}
In present work the conductivity in the basis plane of YBaCuO single crystals slightly doped by Al and Pr with a pre-specified topology of twin boundaries has been investigated. The excess conductivity for the analyzed samples shows dependence like $\Delta\sigma\sim(1-T/T^*)\exp(\Delta_{ab}^*/T)$ in wide temperature range $T_f<T<T^*$, where $T^*$ can be represents as mean field temperature of superconducting transition. The temperature dependence of pseudogap can be satisfactory described in terms of the BCS-BEC crossover theoretical model.
\end{abstract}

\maketitle

\section{Introduction}
Pseudogap anomaly (PA) in high temperature superconductor (HTS)-compounds is an important problem being paid a great interest \cite{Sadovskii01, Prokofyev03, Babaev99}. According to modern conceptions this phenomenon can be a clue to understanding the HTS nature \cite{Prokofyev03, Babaev99}. In the recent literature there are two scenarios for the explanation of the PA origin in HTS-systems. According to the first, PA is related to short-range ordering fluctuations of the "dielectric" type, e.g., antiferomagnetic fluctuations, spin and charge density waves, etc. \cite{Sadovskii01}. The second scenario assumes that Cooper pairs are formed at temperatures much higher than the critical one; $T^* >> T_c$ ($T^*$ is the temperature of opening PA) and further phase coherence realized at $T < T_c$ \cite{Prokofyev03, Babaev99}.

The most convenient objects to study the PA are \ybco compounds. This is due to the possibility of vary substitution of components by isoelectron analogies or by changing of oxygen stoichiometry. It is well known \cite{Dover89, Tarascon88} that doping by aluminum leads to substitution of Cu atoms in the CuO planes while the data about influence of such replacement on transport properties remain quite contradictory \cite{Dover89}. Also interest can be paid to substitution Pr instead of Y. It leads to the suppression of superconductivity while the lattice parameters and oxygen stoichiometry remain unchangeable \cite{Kabede91, Sandu04}. Besides, in YBaCuO single crystals are always present planar defects, i.e. twin boundaries (TB). Influence of such defects on the transport properties in normal state is poorly studied because of experimental problems that are connected with the determination of the contribution of such defects. All things considered, in the present work we studied influence of Al and Pr-doping on the longitudinal conductivity of YBaCuO single crystals with high critical temperature $T_c$ and work the system of unidirectional TB. To minimize the influence of the TB transport current was oriented $\textbf{I}\parallel$TB.
\section{Experiment details and samples}
YBa$_2$Cu$_3$O$_{7-\delta}$, YBa$_2$Cu$_{3-y}$Al$_y$O$_{7-\delta}$ and Y$_{1-z}$Pr$_z$Ba$_2$Cu$_3$O$_{7-\delta}$ single crystals were grown by self-flux method \cite{Vovk07}. Technology of the sample growth, perform of the transport measurements as well as analysis of the transport properties both in normal and in superconductivity states are described in detail in \cite{Vovk07}. Geometry of the experiment is shown in Fig.~\ref{fig:1}, inset (a). Resistive parameters of samples are presented in the TABLE 1.
\begin{table}[h]
\begin{tabular}{|l|l|l|l|l|l|l|} \hline
Sample & $T_c$, K & $\rho_{ab}$(300), $\mu\Omega\cdot$cm & $T^*$, K & $\Delta_{max}^*$, mev & $t^*$\\ \hline
K1 & 91.74 & 155 & 143 & 88.4 & 0.5158\\ \hline
K2 & 92.05 & 421 & 199 & 58.1 & 1.129\\ \hline
K3 & 85.8 & 255 & 110& 98.1 & 0.2099\\ \hline
\end{tabular}
\caption{Resistive parameters for samples \ybco (K1), YBa$_2$Cu$_{3-y}$Al$_y$O$_{7-\delta}$ (K2) and Y$_{1-z}$Pr$_z$Ba$_2$Cu$_3$O$_{7-\delta}$ (K3)}
\end{table}
With the use of published data, regarding the dependence of $Ò_ñ$ from the Al \cite{Dover89, Tarascon88} and Pr \cite{Kabede91, Sandu04} concentrations, we can conclude that Al and Pr concentrations in the K2 and K3 crystal is less that 5~$\symbol{37}$, when the oxygen concentration is $\delta\leq$~0.1 \cite{Vovk07}.
\section{Results and discussions}
\begin{figure}
\includegraphics[clip=true,width=3.2in]{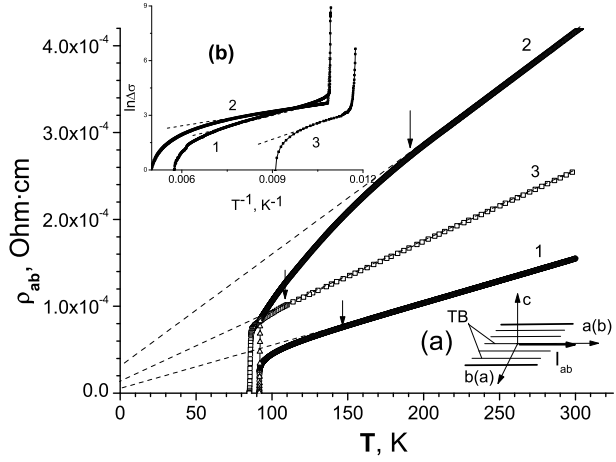}
\caption{\label{fig:1} Temperature dependences of resistance $\rho_{ab}(T)$ of K1 (curve 1), K2 (curve 2) and K3 (curve 3) single crystals. Dashed lines show the extrapolation of linear parts to the zero temperature. Arrows show transition temperatures $T^*$ into pseudogap (PG) regime. Inset (a) shows the schematic geometry of the experiment. Inset (b) shows temperature dependences of the excess conductivity for the same samples in ln$\Delta\sigma$ vs. $1/T$ coordinates. Dashed lines in inset (b) show the approximation of experimental curves by equation~\ref{eq:2}.}
\end{figure}
Fig.~\ref{fig:1} shows that temperature dependences have wide linear part at relatively high temperatures. According to the NAFL theory \cite{Stojkovic97} this is a feature of the normal state. At the temperatures below the characteristic value $T^*$ dependences $\rho_{ab}(T)$ change from the linear law. This is caused by the appearance of an excess conductivity that corresponds to the transition to PG regime \cite{Sadovskii01, Prokofyev03, Babaev99}. As one can see from Fig.~\ref{fig:1} for Y$_{1-z}$Pr$_z$Ba$_2$Cu$_3$O$_{7-\delta}$ sample the linear part of $\rho_{ab}(T)$ is more wide compared to the \ybco but the temperature $T^*$ decreases. It means that temperature interval of excess conductivity becomes narrower. It is significant up to the present article in the Pr-doped YBaCuO compounds at $z\geq$ 0.2 $T^*$  was higher than in undoped samples \cite{Sandu04}. On the other hand, regarding the Al doped crystal, the linear section of the $\rho_{ab}(T)$ dependence is significantly narrower as compared with the dopant-free crystal, and the temperature $T^*$ is displaced, in the section of high temperatures, by more than 55~K. This indicates the relevant expansion of the temperature interval of the excess conductivity existence.

Temperature dependence of an excess conductivity is defined by the equation:
\begin{eqnarray}
{\Delta\sigma = \sigma-\sigma_0},
\label{eq:1}
\end{eqnarray}
where $\sigma_0=\rho_0^{-1}=(A+BT)^{-1}$ is the conductivity obtained from the extrapolation of the linear part to the zero temperature, and $\sigma=\rho^{-1}$ is the experimentally defined value of conductivity in the normal state.

Received experimental dependences $\Delta\sigma(T)$ are shown in inset (b) Fig.~\ref{fig:1} in ln$\Delta\sigma$ vs. $1/T$ coordinates. As we can see, these dependences have linear parts in a wide temperature range. It is possible to describe these features by
\begin{eqnarray}
{\Delta\sigma\propto \exp\left(\frac{\Delta^*}{T}\right)},
\label{eq:2}
\end{eqnarray}
where $\Delta^*$ is the value defines thermal activated process through the energy gap - "pseudogap". Value $\Delta^*$, that is calculated from \ref{eq:2} for K1, K2, K3 samples, is given in TABLE 1.

One can see that doping by aluminum causes significant decrease of the absolute value $\Delta_{K1}^*/\Delta_{K2}^*\approx$ 1.52 while doping by Pr leads to increase of the absolute value $\Delta_{K1}^*/\Delta_{K3}^*\approx$ 0.9.

\begin{figure}
\includegraphics[clip=true,width=3.2in]{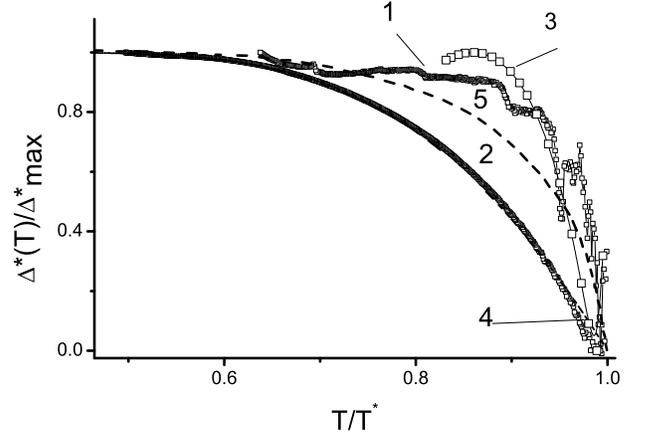}
\caption{\label{fig:2} Temperature dependences of the pseudogap in coordinates $\Delta^*(T)/\Delta_{max}^*$ vs. $T/T^*$ ($\Delta_{max}^*$ is the value of $\Delta^*$ on plateau far away from the $T^*$). The numbering of the curves in the figure is consistent to the Fig.~\ref{fig:1}. Dashed lines (curve 4) and (curve 5) show the dependence $\Delta^*(T)/\Delta(0)$ is calculated according to \cite{Babaev99}.}
\end{figure}

Earlier, the exponential dependence of $\Delta\sigma(T)$ was observed in the YBaCuO film samples \cite{Prokofyev03}. As it was shown in \cite{Prokofyev03}, approximation of experimental data can be much more expanded by entering factor $(1-T/T^*)$. In this case the excess conductivity becomes proportional to the superconducting carrier density $n_s\sim(1-T/T^*)$ and inversely proportional to the pairs that are destroyed by thermal motion:
\begin{eqnarray}
{\Delta\sigma\propto \left(1-\frac{T}{T^*}\right)\exp\left(\frac{\Delta^*}{T}\right)},
\label{eq:3}
\end{eqnarray}
where $T^*$ is the mean field temperature of the superconducting transition. Temperature range $T_c<T<T^*$, in which PA exists, defined by the phase of the order parameter that depends on either oxygen deficiency or doping element concentration. Thus, using the method that was proposed in \cite{Prokofyev03} according to experimental curve ln($\Delta\sigma(T)$)  one can obtain temperature dependence $\Delta^*(T)$ up to $T^*$. Temperature dependences of the pseudogap that were obtained for single crystals K1, K2, K3 are shown in coordinates $\Delta^*(T)/\Delta_{max}^*$ vs. $T/T^*$ in Fig.~\ref{fig:2} by curves 1, 2 and 3 respectively.

In the theoretical work \cite{Babaev99} temperature dependences of the pseudogap were obtained in the mean field approximation in the context of the crossover theory of BCS-BEC for the case of weak \ref{eq:4} and strong \ref{eq:5} coupling:
\begin{eqnarray}
{\Delta(T)=\Delta(0)-\Delta(0)\sqrt{2\pi\delta(0)T}\exp\left(-\frac{\Delta(0)}{T}\right)},
\label{eq:4}
\end{eqnarray}
\begin{eqnarray}
{\Delta(T)=\Delta(0)-\frac{8}{\sqrt{\pi}}\sqrt{-x_0}\left(\frac{\Delta(0)}{T}\right)^{3/2}\exp\left(-\frac{\sqrt{\mu^2+\Delta^2(0)}}{T}\right)},
\label{eq:5}
\end{eqnarray}
 where $x_0=\mu/\Delta(0)$ ( $\mu$ is the chemical potential of the carrier system, $\Delta(0)$ is the value of energy gap at the zero temperature).
Dependences $\Delta^*(T)/\Delta(0)$ vs. $T/T^*$ that were calculated according to \ref{eq:4} and \ref{eq:5} for the value of crossover parameter $\mu/\Delta(0)=$ 10 (limit of the BCS) and $\mu/\Delta(0)=$ -10 (limit of  the BEC) are shown in fig.~\ref{fig:2} by dashed lines 4 and 5, respectively.
One can see that in the case of optimal oxygen doping YBa$_2$Cu$_3$O$_{7-\delta}$ sample, temperature dependence of the pseudogap shows some disagreement with the theory \cite{Babaev99}. The same behavior was observed in YBaCuO film samples \cite{Prokofyev03}. At the same time, for YBa$_2$Cu$_{3-y}$Al$_y$O$_{7-\delta}$ single crystals taking into account some ambiguity in determination of the value $T^*$ accordance with theory is quite satisfactorily.

As one can see in Fig.~\ref{fig:1}, inset (b) the closer to $T_c$ the greater value $\Delta\sigma$. It is known from the theory \cite{Aslamazov68} that in the vicinity of $T_c$ an excess conductivity is caused by the fluctuation carrier coupling processes (FC). If determine the temperature of transition from PG to FC-regime $T_f$ at the point where ln$\Delta\sigma$ vs. $1/T$ deviate up from linear dependence \cite{Babaev99}, one can estimate relative range of PG existence as $t^*=(T^*-T_f)/T_f$. Calculated results show that at small Al-doping (up to $y\approx 0.05$) pseudogap temperature range increases more than two times, from $t^*$ = 0.5158 to 1.1293. On the other hand, the Pr-doped sample with the same concentration results in the inverse effect that is the narrowing of the pseudogap temperature range to 0.2099.
\section{Conclusions}
In conclusion, the excess conductivity $\Delta\sigma(T)$ for the \ybco, YBa$_2$Cu$_{3-y}$Al$_y$O$_{7-\delta}$ and Y$_{1-z}$Pr$_z$Ba$_2$Cu$_3$O$_{7-\delta}$ single crystals shows exponential dependence in wide temperature range $T_f<T<T^*$. In addition, the ratio $\Delta\sigma\sim \left(1-T/T^*\right)\exp(\Delta^*/T)$ can be interpreted in the mean field theory, where $T^*$ represents as mean field temperature of superconducting transition and temperature dependence of the pseudogap can be described satisfactorily in terms of the crossover BCS-BEC theory. Al-doped YBaCuO single crystal shows the wider temperature range of PA realization. In Pr-doped crystal ($z\approx$ 0.05) occurs unusual decreasing of the pseudogap temperature range.



\end{document}